\documentclass[10pt,twocolumn,letterpaper]{article}

\usepackage{cvpr}              %
\usepackage{amsmath,amsfonts}
\usepackage{algorithmic}
\usepackage{algorithm}
\usepackage{array}
\usepackage{textcomp}
\usepackage{stfloats}
\usepackage{url}
\usepackage{verbatim}
\usepackage{graphicx}
\usepackage{xspace}
\usepackage{booktabs} 
\usepackage{multirow} 
\usepackage{amsmath}
\usepackage{xcolor}
\usepackage{colortbl}
\usepackage{tabularx}
\usepackage{makecell}
\usepackage{soul}
\usepackage{ulem}
\usepackage[accsupp]{axessibility}

\newcommand{\latentset}{ \mathbf{S} }

\newcommand{\points}{ \mathbf{P} }
\newcommand{\numpoints}{ N }
\newcommand{\camera}{ \mathbf{\pi} }

\newcommand{\viewnormal}{ n }

\newcommand{\name}{CraftsMan3D\xspace}

\definecolor{cvprblue}{rgb}{0.21,0.49,0.74}
\usepackage[pagebackref,breaklinks,colorlinks,allcolors=cvprblue]{hyperref}

\def\paperID{3686} %
\def\confName{CVPR}
\def\confYear{2025}

\title{CraftsMan3D: High-fidelity Mesh Generation with 3D Native Diffusion and Interactive Geometry Refiner}

\author{Weiyu Li$^{1,2\ast}$,
Jiarui Liu$^{1,2\ast}$,
Hongyu Yan$^{1,2\ast}$,
Rui Chen$^{1}$,
Yixun Liang$^{1}$
\\
Xuelin Chen$^{3}$,
Ping Tan$^{1,2}$,
Xiaoxiao Long$^{1,2\dag}$
\vspace{0.2cm}
\\
{\normalsize $^{1}$HKUST \quad $^{2}$ LightIllusions}
{\normalsize \quad $^{3}$Adobe Research}
   \\
{\normalsize $^\ast${Core contributions} \quad $^\dag${Corresponding author} 
}
}

\begin{document}

\twocolumn[{%
\renewcommand\twocolumn[1][]{#1}%
\maketitle
\begin{center} 
    \centering 
    \includegraphics[width=1.\textwidth]{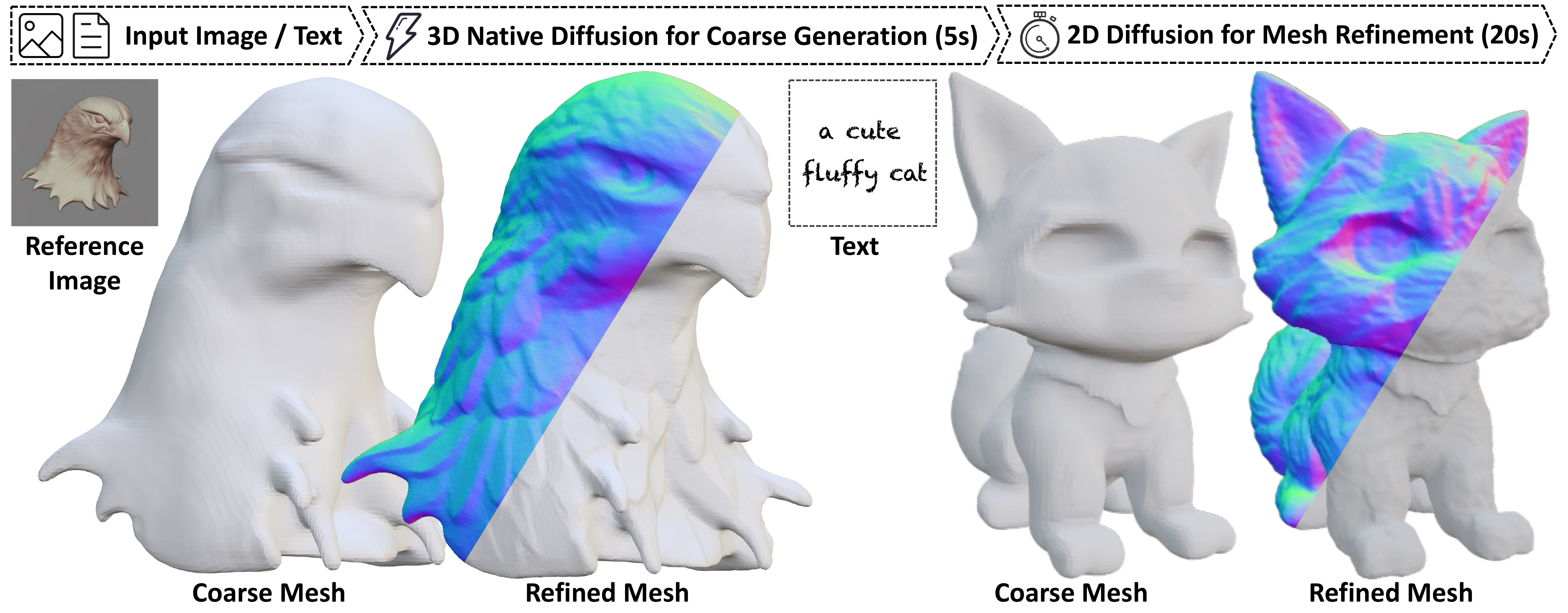} 
    \captionof{figure}{
    Our method, given a single reference image or text prompt, can generate intricate 3D shapes with high fidelity in just 25 seconds. Drawing inspiration from the typical workflow of the craftsman, we start by creating a coarse shape using a 3D native DiT model. We then enhance the surface details using either an automatic global geometry refiner or, more intriguingly, an interactive geometry refiner that allows for users to edit. 
    For more visually compelling results, please refer to the supplementary video.
    } 
    \label{fig:teaser}
\end{center}%
}]

\begin{abstract}

We present a novel generative 3D modeling system, {coined \name}, which can generate high-fidelity 3D geometries with highly varied shapes, detailed surfaces, and, notably, allows for refining the geometry in an interactive manner. 
Despite the significant advancements in 3D generation, existing methods still struggle with 
lengthy optimization processes, 
self-occlusion, irregular mesh topologies, 
and difficulties in {accommodating user editing}, {
consequently impeding their widespread adoption and implementation in 3D modeling softwares.} 
{Our work is inspired by the craftsman, who usually roughs out the holistic figure of the work first and elaborates the surface details subsequently.}
Specifically, we first introduce a robust data preprocessing pipeline that utilizes visibility check and winding mumber to maximize the use of existing 3D data. 
Leveraging this data, we employ a 3D-native DiT model that directly models the distribution of 3D data in latent space, 
generating coarse geometries in seconds. Subsequently, a normal-based geometry refiner enhances local surface details, which can be applied automatically or interactively with user input.
{Extensive experiments} demonstrate that our {method achieves high efficacy in producing} superior quality 3D meshes compared to existing methods.

\end{abstract}
    
\section{Introduction}

The rapid development of industries such as video gaming, augmented reality, and film production has led to a surge in demand for automatic 3D asset creation.
However, existing methods still struggle to produce results that are ready to use.

3D generative methods can be broadly categorized into three types: 
i) Score-Distillation Sampling (SDS) based methods~\cite{poole2022dreamfusion,lin2023magic3d,fantasia3d} typically distill priors in pretrained 2D diffusion models for optimizing a 3D representation, eventually producing 3D assets. 
However, these methods often suffer from time-consuming processing, unstable optimization, and multi-face geometries.
ii) Multi-view (MV) based methods propose generating multi-view consistent images as intermediate representations, 
{from which the final 3D can be reconstructed}~\cite{li2023instant3d,long2023wonder3d,liu2023syncdreamer}. 
While these methods significantly improve generation efficiency and robustness, the resulting 3D assets tend to have artifacts and struggle to generate assets of complex geometric structures.
iii) 3D native generation methods~\cite{nichol2022point,jun2023shap,zhang2024clay,wu2024direct3d} attempt to directly model the probalistic distribution of 3D assets via training on 3D assets. However, due to the limited 3D data and high-dimensional 3D representation, existing 3D generative models can not produce high-fidelity details.
More importantly, all of these methods 
do not support user editing to improve the generated 3D interactively.

Challenges of scaling up native 3D generative models largely due to the uniform requirement of training data. Unlike the standardized structures of text and 2D images, 3D assets are from various sources—procedural functions, 3D modeling, or scanning, resulting in diverse mesh topologies such as closed, open, double-sided, non-manifold that require careful handling to maintain geometric integrity, making uniform dataset creation difficult.
Point-E~\cite{nichol2022point} pioneers a large-scale model trained on millions of 3D assets to generate 3D point clouds from text prompts. While point clouds reduce data acquisition costs, they lack topological detail, limiting their real-world utility.
Implicit distance fields, like signed distance fields (SDF), offer a better alternative due to their continuous, watertight properties, allowing for high-quality 3D mesh extraction. Consequently, existing 3D datasets often require preprocessing to convert meshes into SDFs. Leveraging this, Shape-E~\cite{jun2023shap} improves 3D generation quality, while recent models like CLAY~\cite{zhang2024clay} and Direct3D~\cite{wu2024direct3d} adopt advanced diffusion techniques. However, none of these methods can generate high-fidelity geometric details and limitations in mesh-to-SDF conversions still result in training difficulty.

To tackle problems mentioned above, we first propose an efficient and robust mesh-to-SDF algorithm that maximizes the utilization of existing 3D data~\cite{objaverse, objaverseXL}. By integrating visibility checks with winding number analysis, we significantly enhance the success rate of the watertight conversion and form a high-quality 3D dataset based on Objaverse~\cite{objaverse}.
Built on the 3D data, 
{we present a two-stage generative 3D native generation system, coined CraftsMan, which takes as input single images as reference or text prompts and generates high-fidelity 3D geometries featuring highly varied shapes, regular mesh topologies, and detailed surfaces, and, notably, allows for interactively refining the geometry.}
{Drawing inspiration from craftsmen, who typically begin by shaping the overall form of their work before subsequently refining the surface details,}
{our system is comprised of two stages}:
1) 
{a native 3D diffusion model, that is conditioned on single image and directly generates coarse 3D geometries;} 
and 
2) a robust generative geometry refiner that provides intricate details powered by Poisson Normal Blending and Relative Laplacian Smoothing regularization.

In summary, our main contribution lies in three aspects:

\begin{itemize}

\item A robust and efficient data pre-processing pipeline that integrates visibility checks enhanced by the winding number and significantly improves the success rate of watertight mesh conversion.

\item A simple yet effective 3D Native DiT model. 
{Extensive experiments} demonstrate that our {simple structure achieves high efficacy in producing} superior quality 3D assets compared to existing methods.

\item A novel normal-based interactive mesh refiner which can produce highly enhanced geometries within just 20 seconds and support interactive manipulation, enhancing the generated coarse geometries to better align with the users' envisioned designs.

\end{itemize}

\section{Related work}

In this section, we will first provide a brief review of the relevant literature on
3D generation, followed by a discussion of recent works focused on 
3D native generative models.

\subsection{3D Generation using 2D Supervision}
In recent years, generative models have achieved significant success in producing high-fidelity and diverse 2D images,
and we have seen a surge of interest in lifting this powerful 2D prior to 3D generation.
Most of these methods generate 3D contents, typically in the form of NeRF~\cite{mildenhall2020nerf} or Triplane~\cite{Chan2021eg3d} representations, 
which are turned into images by a differentiable renderer.
Then the multiview images can be compared with either real-world dataset samples or images rendered from 3D models to train a generative model.
\cite{schwarz2020graf,niemeyer2021giraffe,chan2021pi,gao2022get3d} perform GAN-like~\cite{goodfellow2014generative} structure to synthesize 3D-aware images via adversarial training.

However, these methods are often trained on limited views with specific categories, and therefore shows poor generalization on unseen categories.
\cite{poole2022dreamfusion} develop techniques to distill 3D information from a large-scale pretrained 2D text-to-image diffusion models to optimize 3D representation, thus yielding 3D assets. 
Subsequent works~\cite{wang2023prolificdreamer,EnVision2023luciddreamer,fantasia3d,lin2023magic3d,shi2023MVDream,sweetdreamer} are proposed to further enhance the quality of 3D generation.
By leveraging existing powerful 2D priors, these per-shape optimization methods take dozens of minutes and require a huge computational cost.

\begin{figure}[t!]
  \centering
  \includegraphics[width=\linewidth]{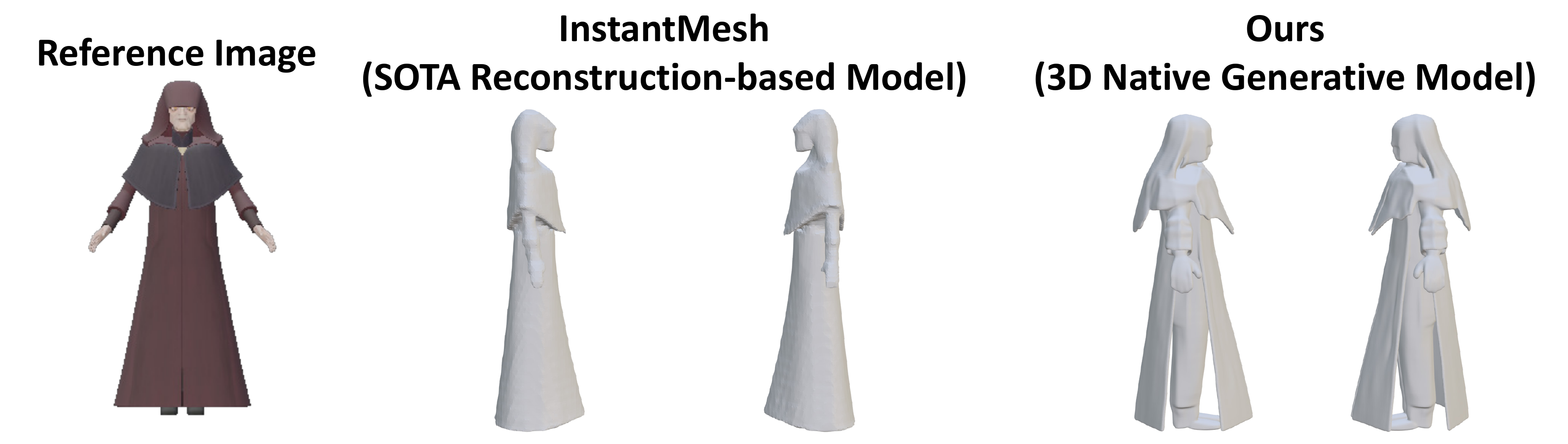}
  \caption{
    Compared to the SOTA reconstruction-based models, our result produces accurate complex geometric structures, including those that are self-occluded in the input images.}.
  \vspace{-5mm}
  \label{fig:disscussion_advantage}
\end{figure}

Instead of performing a time-consuming optimization, recent works ~\cite{long2023wonder3d,li2023instant3d,liu2023one2345,liu2023syncdreamer} attempt to generate multi-view images simultaneously and bring 3D-awareness by finetuning the 2D diffusion models.
The generated multi-view images are then used to reconstruct a 3D shape using sparse view reconstruction algorithms or Large Reconstruction Models (LRM).
Although these methods achieve high efficiency, the generated results are heavily dependent on the quality of the 2D images.
Indirect modeling of 3D probability distributions is insufficient for faithfully recovering geometric information.
Self-occlusion, complex lighting conditions, and multi-view inconsistency are still challenging, usually result in degraded final generation quality, which can be validate in Figure~\ref{fig:disscussion_advantage}.
In contrast, our approach modeling the distribution of 3D data, enabling high-quality mesh generation even with complex inputs.

\subsection{3D Native Generative Models}

Unlike approaches that rely on 2D supervision, many works adopt various 3D representations such as point clouds~\cite{li2018point,zhou20213d,pointflow}, meshes~\cite{nash2020polygen,Liu2023MeshDiffusion}, and implicit functions~\cite{chen2019learning,park2019deepsdf,sun2024recent} to train native 3D generative models. Building up on recently advanced diffusion models~\cite{ho2020denoising}, a series of works began to conduct 3D diffusion models with the representation of point cloud~\cite{luo2021diffusion,zhou20213d}, meshes~\cite{Liu2023MeshDiffusion} and implicit fields~\cite{chou2023diffusion,shue20233d,yariv2023mosaicsdf}.
However, training these 3D generative models directly on 3D data is quite challenging,
due to the high memory footprint and computational complexity. To tackle these challenges, inspired by the success of latent diffusion~\cite{rombach2022high}, recent studies~\cite{hui2022wavelet, zhao2023michelangelo} first compress 3D shapes into compact latent space, and then perform diffusion process in the latent space. For instance,
\cite{zhang2022dilg} and ~\cite{shape2vecset} propose a method to encode occupancy fields using a set of either structured or unstructured latent vectors. Neural Wavelet~\cite{hui2022wavelet} advocates a voxel grid structure containing wavelet coefficients of a Truncated Signed Distance Function (TSDF).
One-2-3-45++~\cite{liu2023one2345++} and XCube~\cite{ren2024xcube} focus on explicit dense grid volume. 
The most recent works, Michelangelo~\cite{zhao2023michelangelo} and CLAY~\cite{zhang2024clay}, train a diffusion model on latent set representations, and Direct3D~\cite{wu2024direct3d} explores a triplane representation to enhance training scalability.
However, these works often suffer from lacking geometric details, over-smoothing surfaces, and unstable training processes. Our work harnesses the feed-forward nature of 3D diffusion models while enhancing its generalization capability by leveraging the prior from pre-trained multi-view 2D diffusion as the condition. This approach significantly facilitates zero-shot ability and robust generation.

\section{Method}
\label{sec:mesh_generation}
\begin{figure}[t!]
  \centering
  \includegraphics[width=\linewidth]{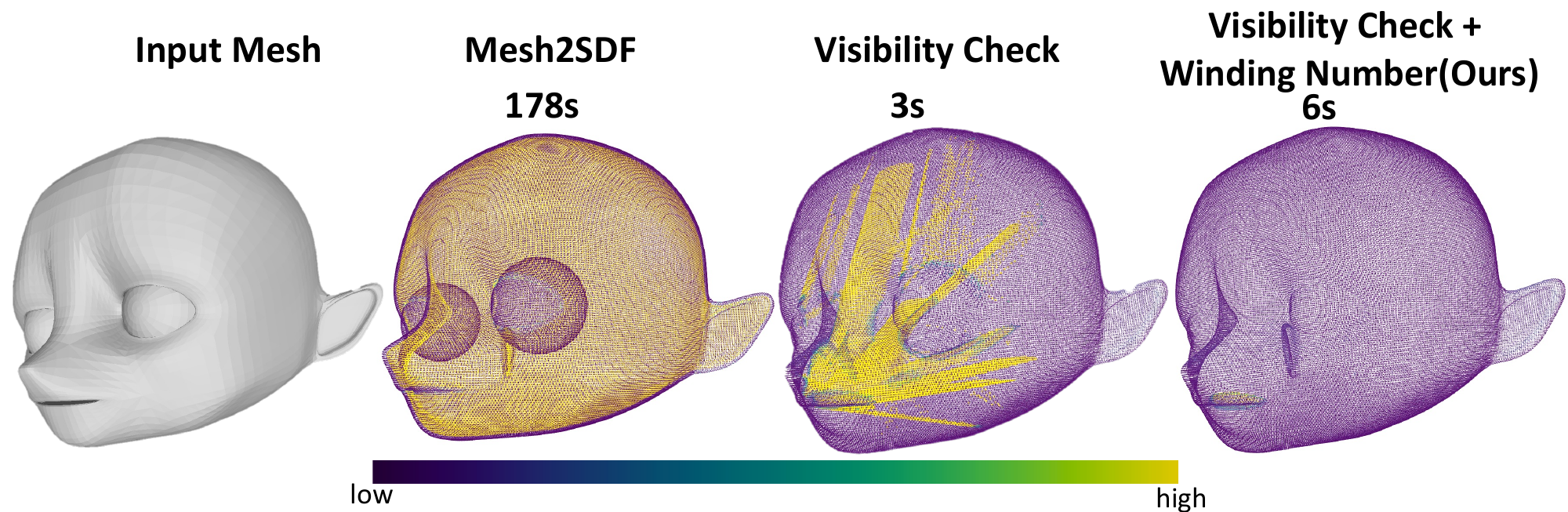}
  \caption{
    Error maps of different mesh-to-sdf methods.  We sample surface points from the processed meshes for each method and show the differences compared to the ground truth mesh.
  }
  \vspace{-5mm}
  \label{fig:data_preprocess}
\end{figure}

\begin{figure*}[t!]
  \centering
  \includegraphics[width=\linewidth]{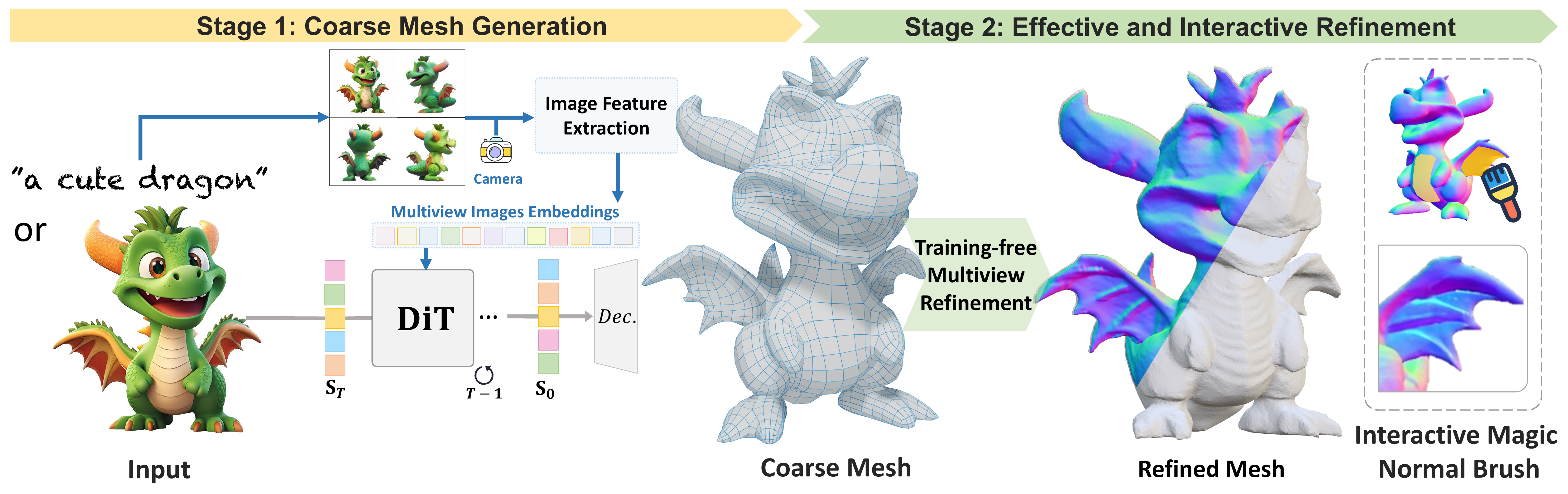}
  \caption{
    Overview of \name. 
    We first using a multi-view diffusion model to generate a multi-view image from the input single image or text prompt.
    The generated multi-view image is then fed into our Latent Set-based DiT model as conditioning to produce a coarse mesh. 
    Finally, a dedicated refinement module is employed to improve or edit the surface normals of the coarse geometry, enhancing with intricate details. 
    In particular, this refinement module features two key usages, namely the automatic global refinement and interactive magic brush, that contribute to efficient and controllable 3D modeling of high-quality meshes.
  }
  \label{fig:overview}
\end{figure*}

Our 3D generation framework mirrors the 3D artist's workflow, which begins typically with the creation of a rough geometry that is then refined in the subsequent stage.
Figure~\ref{fig:overview} illustrates our generative 3D modeling workflow, that is capable of producing high-quality, detailed 3D assets.

In this section, 
We begin by introducing our data preprocessing (Sec.\ref{sec:Data_processing}), which significantly improves the success rate of watertight conversion and maximizes the utilization of existing 3D data. 
Following this, we train a Variational Auto-Encoder (VAE) on the watertight meshes to learn latent set-based representations\cite{shape2vecset} and output a TSDF field. 
Next, we train a dedicated DiT-based denoising network that operates on these learned latent representations, using the intermediate multi-view image as conditioning (Sec.\ref{sec:mv_diffusion}). Finally, our framework features a normal map-based geometry refinement scheme (Sec.\ref{sec:refinement}).

\subsection{Data Preprocessing.}
\label{sec:Data_processing}
Standardizing the geometric data is essential for effectively training a 3D generative model. Due to the significant noise in the geometry and appearance, we first filter out low-quality meshes, including those with point clouds, thin structures, holes, and textureless surfaces to form our initial subset. 
Ensuring that the mesh is watertight is also essential for extracting the SDF (Signed Distance Function) field from the processed meshes as supervision~\cite{zhang2024clay} when training a Shape VAE~\cite{shape2vecset, zhao2023michelangelo}.
Although the dataset proposed in ~\cite{objaverse, objaverseXL} claims to have nearly ten million objects, the vast majority of it is non-watertight, such as scanned point clouds and planes, resulting in less than 1\% of the data can be directed used. Therefore, we propose an efficient and effective method for converting mesh into a watertight one.

\paragraph{Winding Number-Enhanced Watertight Conversion.} 
Dual Octree Graph Networks (DOGN)~\cite{dualocnn} proposed a "mesh-to-SDF" approach, which requires a significant amount of time. 
CLAY~\cite{zhang2024clay} introduced a "visibility check" method for remeshing, which maximizes positive volume while faithfully preserving geometric features. However, as shown in Figure~\ref{fig:data_preprocess}, for the non-manifold objects with holes, it is easy to encounter floaters inside the converted mesh. 
To tackle these challenges, we enhance the visibility check by incorporating the concept of the winding number~\cite{windingnum}, which is an effective tool for determining whether points are inside or outside a shape. When the input point cloud has well-defined normals, the winding number can reliably differentiate between the inside and outside in a global manner. 
Specifically, we first randomly choose 50 cameras on a sphere and use a dense grid with a resolution of 256 or 512 for visibility check. If the center of a grid cell is not visible to all the cameras, we further check the winding number of it. Once the value of the winding number indicator function is greater than a threshold, which we set to 0.75 by default, we treat that point as being inside the object. Thus, we can get robust inside-outside test results.
This approach statistically improves our watertight conversion success rate from 60\% to 80\% on ~\cite{objaverse}. Please refer to the supplementary for more details.

\subsection{Multi-view guided 3D generation model}
\label{sec:3D_native_diffusion}

\begin{figure}[t!]
  \centering
  \includegraphics[width=\linewidth]{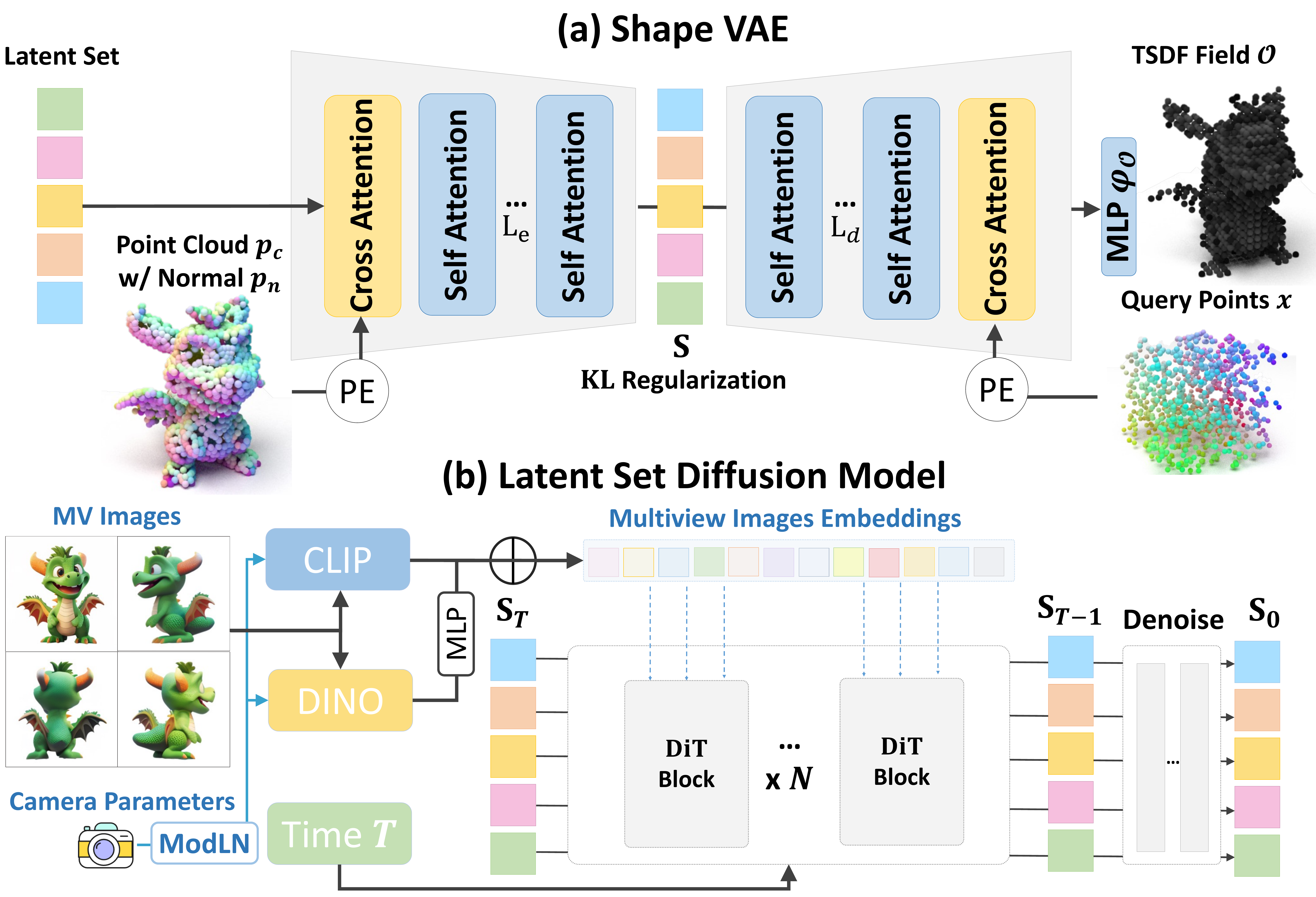}
  \caption{
    The illustration of 3D generation. (a.) We first train a 3D Variational Autoencoder (VAE) to compress 3D shape into a latent space, which takes point clouds with normals as input and outputs TSDF fields. (b.) With the learned latent space, we train a 3D Latent Set DiT Model that using multi-view images as conditions.
  }
  \label{fig:3dnativegeneration}
\end{figure}

\paragraph{3D Shape VAE.} 
\label{sec:shape_vae}
Following~\cite{zhao2023michelangelo}, 
we adopt a Perceiver-based~\cite{jaegle2021perceiver} shape VAE to encode the 3D shape into a set of latent vectors $\latentset$ and then decode them to reconstruct the neural field of the original 3D shape.
Figure~\ref{fig:3dnativegeneration}(a) shows the network architecture.
Specifically, for each 3D shape, we first sample on the 3D surface to obtain a set of points $\points_{c}\in\mathbb{R}^{\numpoints\times3}$, as well as a set of surface normal vectors $\points_{n}\in\mathbb{R}^{\numpoints\times3}$ at these point positions.
The encoder is trained to map points $\mathbf{P}_c$ and $\mathbf{P}_n$ into a latent vector set $\mathbf{Z}$, which a decoder then translates into an implicit field representation. Notably, we replace the original occupancy field with a TSDF field using a threshold of $1/256$ for stable optimization and better performance.

\paragraph{Multi-view Guided 3D Diffusion Model.}
\label{sec:mv_diffusion}
Instead of directly using the input single image or text prompt as conditioning, 
our DiT-based diffusion model is conditioned on the multi-view (MV) images that capture the target 3D asset.
During inference the pre-trained text-based~\cite{shi2023MVDream} or image-based~\cite{wang2024crm, long2023wonder3d, li2024era3d} MV diffusion models are used to generate the corresponding MV image from the input single image or text prompt accordingly.
MV images generated by recent MV diffusion models offer richer geometric and contextual information compared to using a single image or text alone.
As a result, the multi-view conditioned DiT model enables improved generation of various 3D shapes, particularly on unobserved regions from the single input image.

With the latent set representation $\latentset$ of a shape and its corresponding multi-view images $\hat{\mathbf{y}}$,
we now train a MV-conditioned DiT model.
To make image embeddings be aware of the camera position, we follow the method ~\cite{li2023instant3d} to modulate the camera parameters $\camera$ during the feature extraction, by employing an adaptive layer normalization (adaLN)~\cite{perez2018film}.
Formally, the conditioned embeddings $c$ can be represented by: 
\begin{equation}
\small
  c = \varphi_{clip}(\hat{\mathbf{y}}, ModLN(\pi)) + \varphi_{mlp}(\varphi_{dino-v2}(\hat{\mathbf{y}}, ModLN(\pi)))
\end{equation}
where $\varphi_{clip}$ and $\varphi_{dino-v2}$ are pretrained CLIP~\cite{radford2021learning} and DINO-v2~\cite{caron2021dino} and $\varphi_{mlp}$ is a small MLP that aligns DINO features with CLIP features.
Then, we can learn the Multi-view guided Latent Set Diffusion Model (LSDM) via:
\begin{equation}
\begin{aligned}
  \mathcal{L}_{LSDM} :=\mathbb{E}_{\mathcal{E}(x), y, \epsilon \sim \mathcal{N}(0,1), t} \left[\left\|\epsilon-\epsilon_\theta\left(\latentset_t, t, c\right)\right\|_2^2\right],
\end{aligned}
\end{equation}
where $\epsilon_\theta$ is build on a DiT~\cite{Peebles_DiT} model, 
$t$ is time step and $\latentset_t$ is a noisy version of $\latentset_0$.
To reduce the number of parameters and computational cost, we employ adaLN-single~\cite{chen2023pixartalpha} in each DiT block.

\subsection{Normal-based Geometry Refinement}
\label{sec:refinement}

To further enhance the coarse mesh, we propose to improve the initial mesh using normal maps as an intermediate representation.
We first render the normal maps of coarse mesh and then leverage normal-based diffusion to enhance the rendered normals with intricate details.
Subsequently, the refined normals serve as supervision to optimize the mesh, thus yielding a refined mesh with rich details.
Moreover, this process also can be performed in an interactive way. Users can select the areas to be edited using a painting brush, creating a binary mask that indicates the regions to be updated. Please refer to the supplementary video for more visual results.

\paragraph{Intermediate Normal Guidance Generation}
We adopt ControlNet-Tile~\cite{zhang2023adding} that is finetuned on a normal dataset~\cite{objaverse, huang2023humannorm} to enhance the rendered normals with details.
A pivotal challenge arises from the inconsistencies observed in the normal images generated by diffusion models across different views. 
Recent advancements, as detailed in \cite{shi2023MVDream, long2023wonder3d}, address this issue by employing a cross-view attention mechanism. 
Interestingly, we have observed that the cross-view attention mechanism can be directly applied to our task in a training-free manner. 
This is partially attributable to the inherent constraints of the coarse normal maps and the design of ControlNet-Tile, which hallucinates new details without significantly altering the original input conditions. 
Formally, during the diffuse process, for the $i_{th}$ view with a rendered normal map $\viewnormal_i$, we replace the $K$ and $V$ in the original attention layer with:
\begin{equation}
\label{eq:sync_attn}
K=W^K\left[z^{0},\cdots, z^{K}\right], V=W^V\left[z^{0},\cdots, z^{K}\right],
\end{equation}
Here, the key $K$ and value $V$ are globally shared for all input views.

\begin{figure}[t!]
  \centering
  \includegraphics[width=\linewidth]{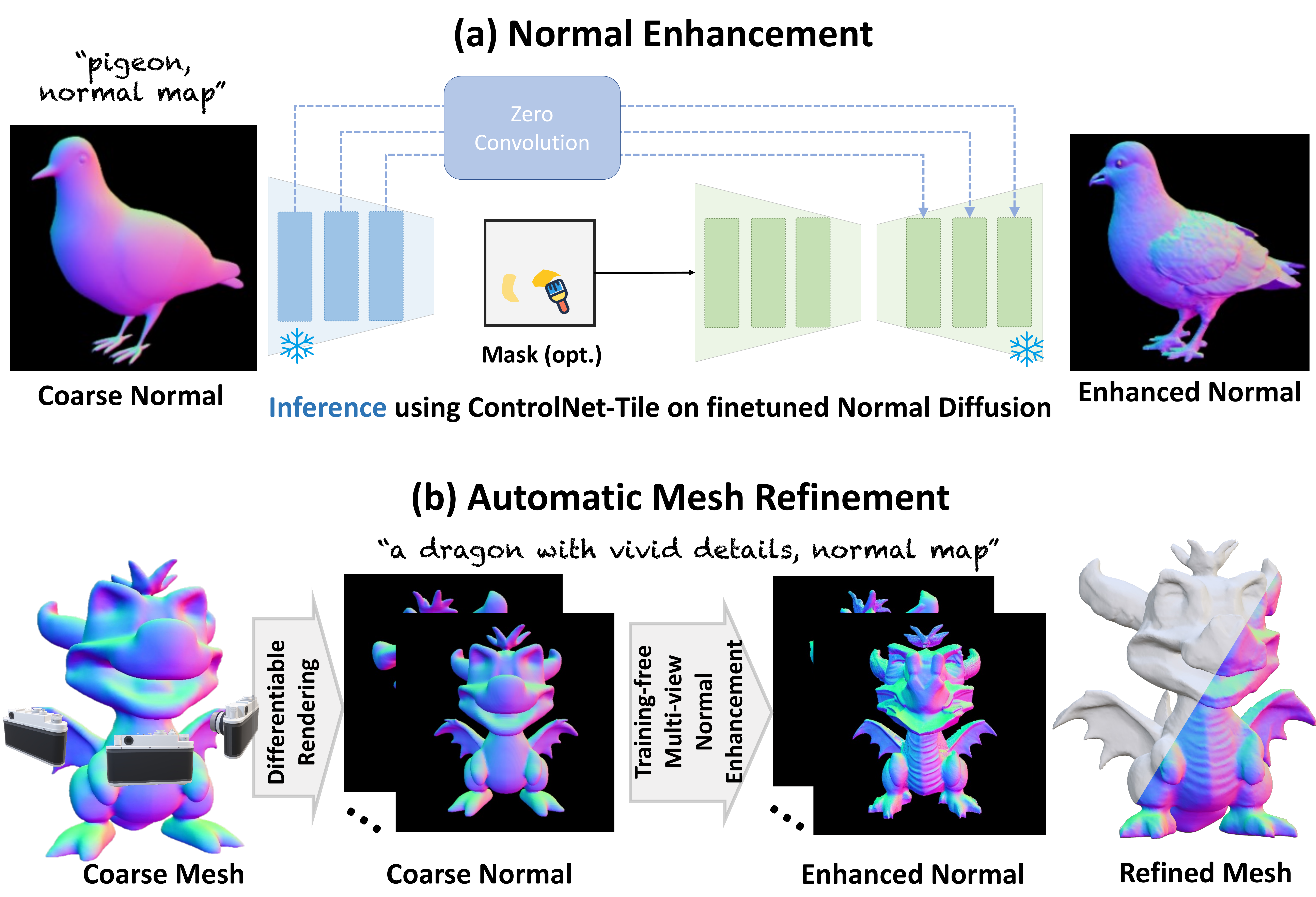}
  \caption{
  The illustration of surface normal-based geometry refinement. (a) The normal-adapted diffusion model is combined with ControlNet-Tile to enhance a normal with intricate details. (b) The automatic mesh refinement process via training-free cross-view attention. %
  }
  \label{fig:mesh_refine}
\end{figure}

\paragraph{Shape Editing via Normal-based Optimization}
We advocate for direct vertex optimization through continuous remeshing~\cite{palfinger2022continuous}, 
which is favored for its computational efficiency and explicit control over the optimization process. 
Given a mesh with vertices $V$ and faces $F$, we optimize the mesh details by directly manipulating the triangle vertices and edges,
with the supervision of the refined normal maps $\hat{n}_{i}$.
Specifically, in each optimization step, we render normal maps from the current mesh via differentiable rendering, denoted as $\mathcal{R}_{n}(V, F, \pi_{i})$.
Then, we minimize the L1 differences between the rendered normals and the refined normals via:
\begin{equation}
\mathcal{L}_{remeshing} = \sum_{i}^{} \| \hat{n}_{i}-\mathcal{R}_{n}(V, F, \pi_{i})  \|_{1}^{1},
\end{equation}
where $\mathcal{R}_{n}$ denotes the differentiable normal rendering function and $\pi_i$ is the camera information of $i_{th}$ rendering camera.
In each step, an update operation is executed to update the position for each vertex according to the gradient computed in the loss backward process.

\begin{figure}[t!]
  \centering
  \includegraphics[width=0.9\linewidth]{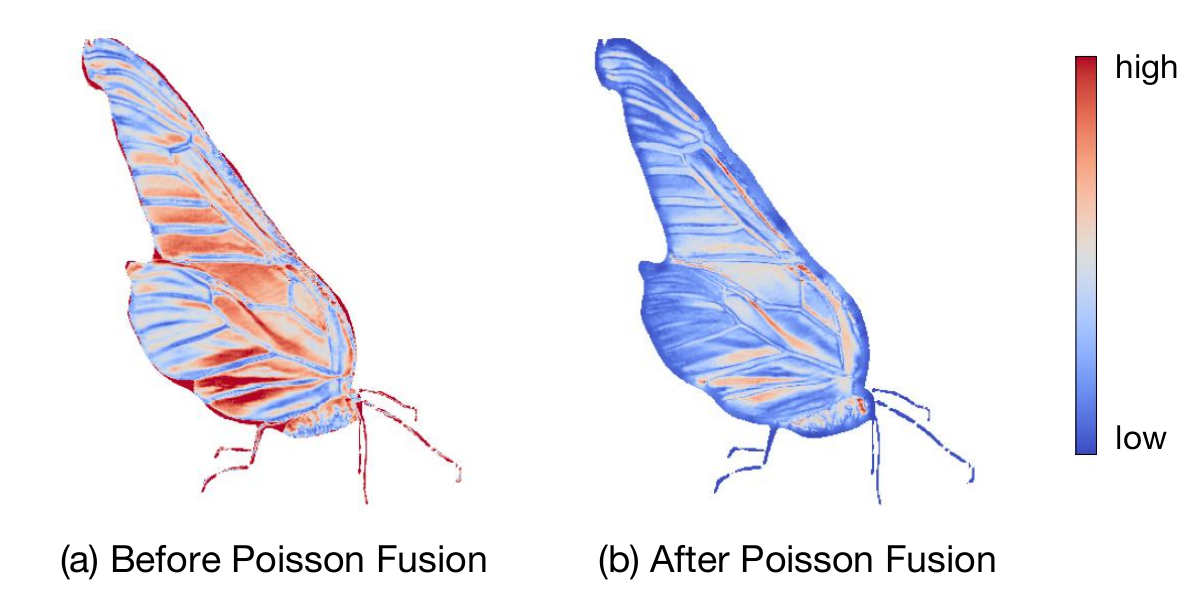}
  \caption{
  Distance map with coarse normal. Normal maps enhanced by stable diffusion contain low-frequency changes from original normal map(shown in red in (a)), which will result in global distortion of input shapes. Applying Poisson Fusion eliminate global distortions, resulting in the preservation of global shapes and the enhancement of high details(b). 
  }
  \label{fig:poisson_fusion}
\end{figure}

\paragraph{Poisson Normal Blending}
Diffusion models generate normals maps by regarding them as a specific domain of images. 
We found that normal maps generated this way sometimes are inaccurate, which results in unstable optimization. 
Figure~\ref{fig:poisson_fusion}(a) shows the pixel-wise L2 distance between the normal map rendered from coarse mesh and the normal map enhanced by the normal stable diffusion, which shows significant changes in the initial shape and leads to stretched shape during the optimization.
To address this, we try to eliminate the influence of those low frequency changes and only take use of the local details contained in the enhanced normal map. We accomplish this by employing the efficient and traditional Poisson Blending algorithm~\cite{perez2023poisson}:
\begin{equation}
n_{fused} = \Gamma\left ( \hat n, \mathcal{R}_{n}(V, F, \pi), m \right ) .
\end{equation}
we denote $\Gamma$ as the Poisson Blending algorithm, $\mathcal{R}_{n}(V, F, \pi)$ and $\hat n$ are the rendered normal map and enhanced normal map respectively. $m$ denotes the mask rendered from coarse mesh, which will be used to label the target region to be fused.

\paragraph{Relative Laplacian Smoothing}
Previous methods~\cite{palfinger2022continuous} often achieve stable optimization by introducing Laplace regularization term. However, this term avoids undesirable results by forcing each vertex close to the coordinate origin in a local Laplace coordinate, which inevitably cause the shrink of the shape. Fortunately, in our detail enhancement task, our initial coarse mesh contains a good prior, thus we do not need to constrain the smoothness by enforcing the Laplace coordinate to zero, but punishing the change of the Laplace coordinate comparing to the initial shape, which is called relative Laplacian smoothing term. 
Given a coarse shape with vertices $\mathbf{x}$, we compute the initial Laplace coordinate by $V^\mathbf{W}_{init}=\mathbf{W_{init}}V_{init}$, here $V_{init}$ is the initial vertex coordinate of coarse mesh, $\mathbf{W_{init}}$ is the corresponding Laplacian matrix. Then in every optimization step, we regularize the deformation process by
\begin{equation}
x\gets x + \lambda v (\mathbf{W} V-V^\mathbf{W}_{init}),
\end{equation}
where $x_{init}$ is the initial vertex position, $\lambda$ is a smoothing hyperparameter. Please refer to the ~\cite{palfinger2022continuous} for more details.

\begin{figure*}[t!]
  \centering
  \includegraphics[width=0.94\linewidth]{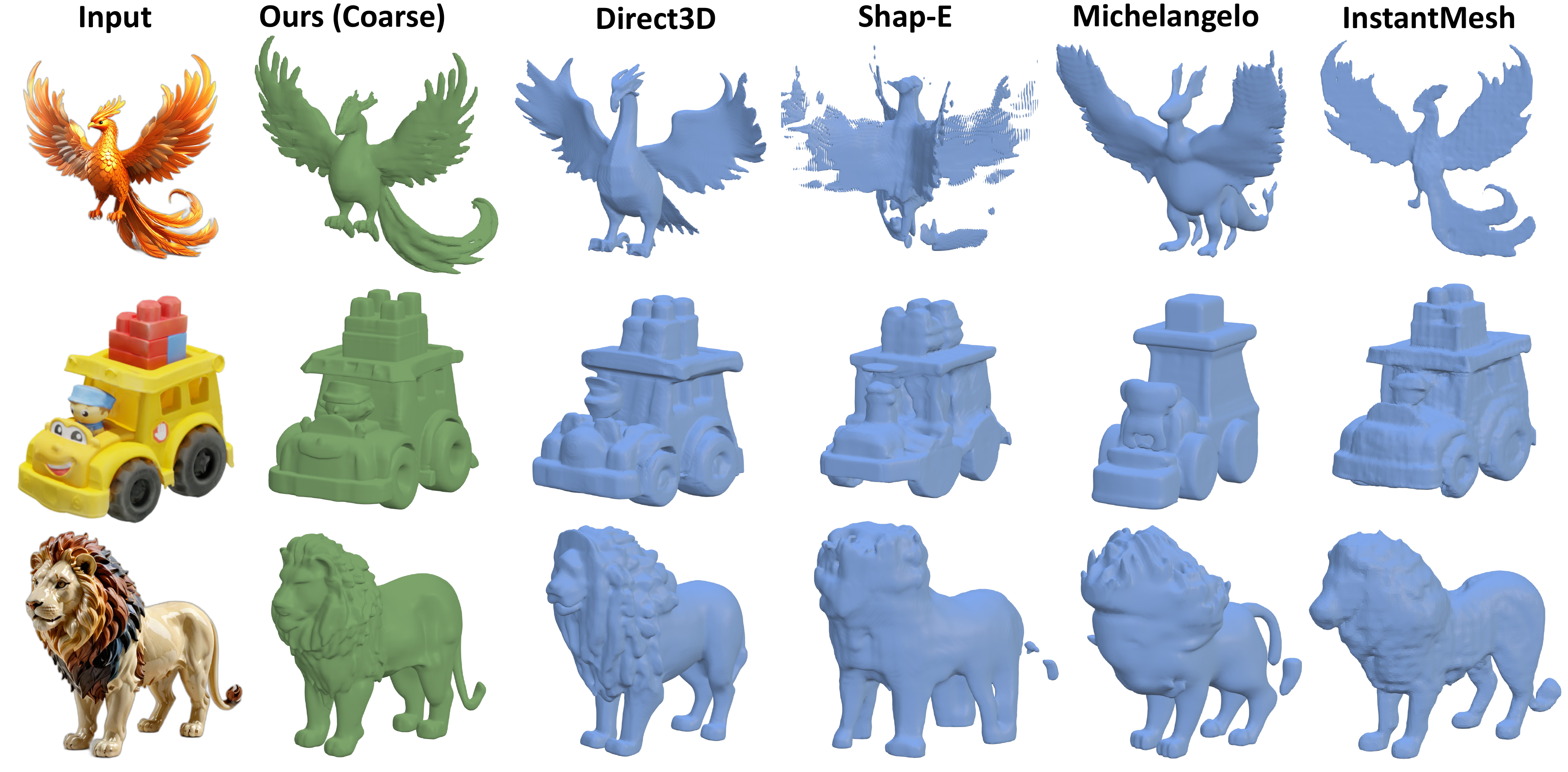}
  \caption{
  Qualitative comparisons with baseline methods for the task of single-view generation. 
    }
  \label{fig:comparisons}
\end{figure*}

\section{Experiments}

\begin{table}[t!]

\caption{
Quantitative comparison with baseline methods on the GSO dataset~\cite{downs2022google}. We follow \cite{liu2023syncdreamer, wang2024crm} and randomly choose 30 shapes from GSO for comparison. Each shape in aligned by conducting an ICP register to calculate the metrics\cite{melaskyriazi2023realfusion}. 
}
\footnotesize
\setlength{\tabcolsep}{4.5pt}
\centering

\begin{tabular}{clccc} 
\toprule
Type  &  Method  &  CD$\downarrow$ & IoU$\uparrow$ & Time$\downarrow$ \\
\midrule
\multirow{3}{*}{\begin{tabular}[c]{@{}c@{}}Recon.-based \\ Model\end{tabular}} & One-2-3-45~\cite{liu2023one2345} & 0.0629 & 0.4086 & \textasciitilde 45s \\
& zero123~\cite{liu2023zero1to3} & 0.0339 & 0.5035 & \textasciitilde 10min \\
& InstantMesh~\cite{xu2024instantmesh} & \underline{\textbf{0.0187}} & \underline{\textbf{0.6353}} & \textasciitilde 10s \\
\midrule
\multirow{2}{*}{\begin{tabular}[c]{@{}c@{}}SDS-based \\ Model\end{tabular}} & Realfusion~\cite{melaskyriazi2023realfusion} & 0.0819 & 0.2741 & \textasciitilde 90min \\
& Magic123~\cite{Magic123} & 0.0516 & 0.4528 & \textasciitilde 60min \\
\midrule
\multirow{6}{*}{\begin{tabular}[c]{@{}c@{}}3D \\ Generative\\ Model\end{tabular}} & Point-E~\cite{nichol2022point} & 0.0426 & 0.2875 & \textasciitilde 40s \\
& Shap-E~\cite{jun2023shap} & 0.0436 & 0.3584 & \textasciitilde 10s \\
& Michelangelo~\cite{zhao2023michelangelo} & 0.0404 & 0.4002 & \textasciitilde 3s \\
& One2345++~\cite{liu2023one2345++} & 0.0437 & 0.3386 & \textasciitilde 20s \\
& Ours & \textbf{0.0291} & \textbf{0.5347} &\textasciitilde 5s \\
\bottomrule
\end{tabular}
\vspace{-3mm}
\label{tab:results}
\end{table}

\begin{table}[t!]

\caption{
Quantitative comparison on subset which contained self-occlusion in the input images.
Our 3D generative model demonstrated a significant performance.
}
\footnotesize
\setlength{\tabcolsep}{13.5pt}
\label{tab:subset}
\centering

\begin{tabular}{lccc} 
\toprule
Method  &  CD$\downarrow$ & IoU$\uparrow$ \\
\midrule
InstantMesh~\cite{xu2024instantmesh} & {0.04909} & {0.50151} \\
Ours & \textbf{0.03943} & \textbf{0.53215} \\
\bottomrule
\end{tabular}
\vspace{-3mm}
\label{tab:results_subset}
\end{table}

To validate the effectiveness of our proposed workflow, we extensively evaluate our proposed framework using a rich variety of inputs.
We present the qualitative and quantitative evaluation of our method as described in Section~\ref{sec:mesh_generation} and Section~\ref{sec:refinement},
as well as comparison results against other baseline methods, showing the effectiveness and efficiency compared to other generation methods.
We also conduct ablation studies to validate the effectiveness of each component in our framework, as described in Section~\ref{sec:ablation}.
More intriguing visual results can be found in our accompanying video and supplementary.
\begin{figure}[t!]
  \centering
  \includegraphics[width=\linewidth]{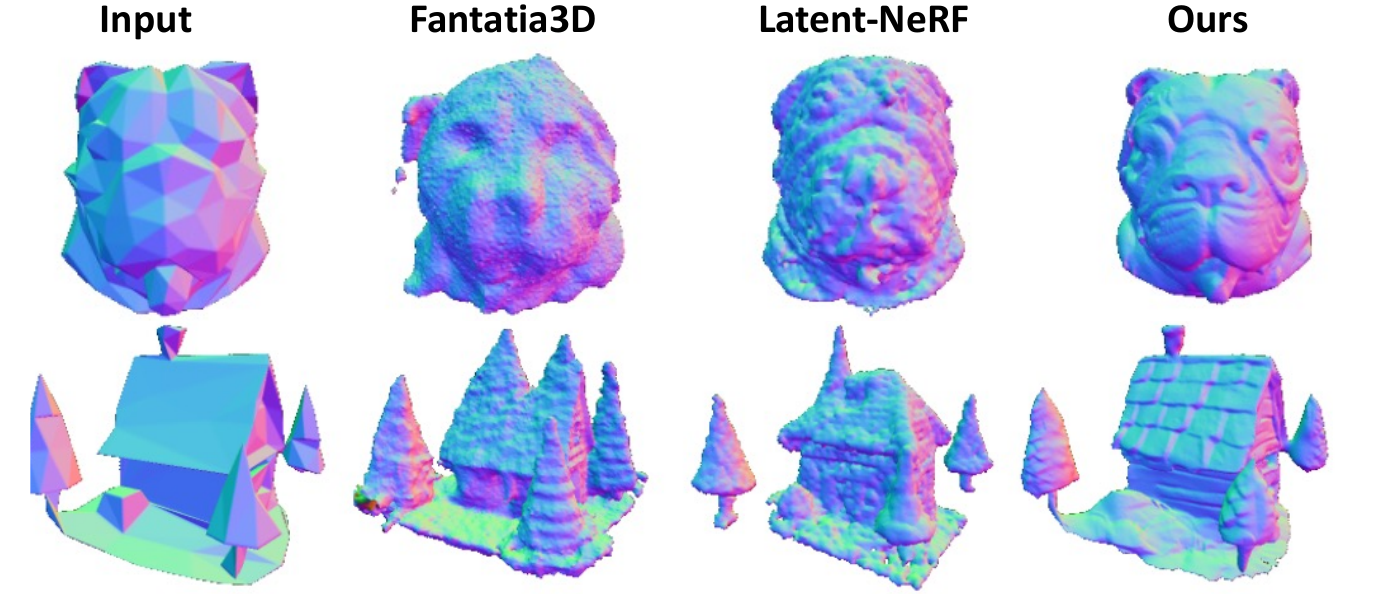}
  \caption{
  Qualitative comparisons with baseline methods for mesh refinement. 
  To better showcase the effects of our mesh refinement, we performed a decimation operation on the input mesh.
  }
  \label{fig:comparisons_mesh_refine}
\vspace{-10px}
\end{figure}

\subsection{Implementation Details}
\label{sec:implementation}

We follow the same architecture as in~\cite{zhao2023michelangelo} for our shape auto-encoder, with the exception of the layer dedicated to contrastive learning, and for our latent set diffusion model.
The shape auto-encoder is based on a perceiver-based transformer architecture with 185M parameters,
while the latent set diffusion model is based on a DiT, comprising 500 million parameters. We train the diffusion model on 32 A800 GPUs using ground truth multi-view images, which share common approaches in related works in this area like \cite{liu2023one2345++, hong2024lrmlargereconstructionmodel}, etc.
Additional details, including dataset, training settings can be found in our supplementary.

\subsection{Evaluation of Mesh Generation}
\label{sec:mesh_generation}

In this evaluation, we focus on presenting the quality of our 3D generation model through a variety of results, and also present quantitative data for reference. We compare our model with several 3D generative models~\cite{jun2023shap,zhao2023michelangelo, wu2024direct3d} and the state-of-the-art large reconstruction model(LRM)~\cite{xu2024instantmesh}.
Given that CLAY~\cite{zhang2024clay} is not publicly available and our request to obtain their results haven't been responded, we only present the visual results in supplementary.

As shown in Fig.~\ref{fig:comparisons}, our 3D native diffusion model produces coarse geometry with regular topology, and the coarse meshes are further enhanced with more intricate details. On the contrary, the 3D native counterpart Shap-E tends to produce noisy surfaces and incomplete shapes, while Michelangelo produces over-smoothed geometries and also suffers from shape ambiguity, like the second example in Fig.~\ref{fig:comparisons}.
Although InstantMesh produces accurate geometries, it can not handle complex geometry structures which results in adhesive geometry and lacks geometric details, take the phoenix in the first line for an example. Compared with Direct3D, our method achieves better consistency between the input image and the generated mesh.

Following the prior works~\cite{liu2023syncdreamer,long2023wonder3d,xu2024instantmesh}, we also employ the Google Scanned Object dataset~\cite{downs2022google}—a rich collection of common everyday objects—to evaluate the performance of our 3D Diffusion Model in generating 3D models from single images. We adopt widely-used Chamfer Distances (CD) and Volume Intersection over Union(IoU) as the metrics. For each object in the evaluation set, we use the front view image as input. 
To align the input for a fair comparison, we first generate multi-view images from input image using existing multi-view diffusion models~\cite{long2023wonder3d,wang2024crm}.
The quantitative evaluation of the quality of our image-to-3D generation is shown in Table~\ref{tab:results}. Our method surpasses all the generation based methods and displays comparable results in a shorter time compared to the reconstruction based method InstantMesh~\cite{xu2024instantmesh}. We notice that the distribution of the GSO dataset is kind of monotonous,lacking mesh with complex structures and self occlusion, which is exactly where our model excels. To fully demonstrate the superiority of our method, we randomly choose a subset from the Objaverse dataset for further evaluation. As shown in Table~\ref{tab:subset}, in this dataset with more complex geometries, the performance of our method is superior to InstantMesh. We also report the time consumption of different methods. In contrast to the SDS-based methods that usually require hours to optimize, our method obtains the resulting mesh in just a few seconds.

\subsection{Evaluation of Mesh Refinement}

\begin{table}[t!]

\caption{
Quantitative comparison for mesh refinement
}
\footnotesize
\setlength{\tabcolsep}{0.5pt}
\centering

\begin{tabular}{lcc} 
\toprule
Method  &  CLIP similarity $\uparrow$ & \quad Time $\downarrow$ \\
\midrule
Fantasia3D~\cite{fantasia3d}  & {0.2567} & \quad  \textasciitilde 15min \\
Latent-NeRF~\cite{metzer2022latentnerf}  & {0.2725} & \quad \textasciitilde 1h \\
Ours & \textbf{0.2821} & \quad \textasciitilde 20s \\
\bottomrule
\end{tabular}
\vspace{-3mm}
\label{tab:results_meshrefine}
\end{table}

To further assess the efficiency of our mesh refinement technique, we compare our method with recent approaches, specifically Fantasia3D~\cite{fantasia3d} and Latent-NeRF~\cite{metzer2022latentnerf}. To reduce the influence of the initial mesh and validate the strong detail enhancement power of our refinement, we reduce the number of face of initial shapes to 500. 
For the comparison with Fantasia3D, we employ the coarse mesh for initialization and only conduct the geometry modeling stage. In the case of Latent-NeRF, we use the input mesh as Sketch Shapes and train the NeRF in Sketch-Shape mode. All comparative experiments were conducted under their default settings. The visual results presented in Figure~\ref{fig:comparisons_mesh_refine} demonstrate that our mesh refinement technique outperforms previous methods, producing not only clear and coherent outcomes but also effectively integrating high-quality details without compromising the overall structural integrity of the original mesh.
Additionally, we provide a quantitative evaluation of our mesh refinement. We selected 20 objects from the Objaverse dataset and employed the same text descriptions as guidance. Table~\ref{tab:results_meshrefine} presents the CLIP~\cite{radford2021learning} similarity scores and the corresponding running times for each method. Our mesh refinement achieved a higher CLIP similarity compared to previous methods, while also demonstrating faster refinement speeds.

\subsection{Ablation Study}
\label{sec:ablation}
We conduct comprehensive ablation studies to substantiate the effectiveness of each design element within our workflow,
showing the importance of each component in the generation of high-quality 3D meshes.

\begin{table}[t!]
\caption{
Ablation study of multi-view guided 3D diffusion model
}
\footnotesize
\setlength{\tabcolsep}{12pt}
\centering

\begin{tabular}{lccc} 
\toprule
Method  & CD$\downarrow$ & IoU$\uparrow$ \\
\midrule
w/o MV condition & 0.0317 & 0.5892 \\
w/o Camera Injection & 0.0249 & 0.6561 \\
ours-Cost Volume & 0.0223 & 0.6583 \\
ours & \textbf{0.0188} & \textbf{0.7059} \\
\bottomrule
\end{tabular}

\vspace{-10px}
\label{tab:ablation}
\end{table}

\textit{Mutil-view images condition.}
In comparison to the single-image condition, the multi-view images generated by the 2D diffusion model offer enhanced information about the object, which is advantageous for generating unseen parts of 3D meshes. By incorporating camera poses into the image feature extractor, our model can better differentiate embeddings from various views of the object, ultimately leading to more accurate 3D shape generation. In the absence of camera pose information, the model is prone to producing 3D geometries with incorrect orientations. Unlike CLAY~\cite{zhang2024clay}, which employs a cost volume that integrates camera pose information before feeding it into their diffusion model, which requires precise camera poses for accurate back projection. To demonstrate the superiority of our design in the context of multi-view images with camera pose injection, we conducted a comparison on our selected subset, which evaluated by the metrics of Chamfer Distance (CD) and Intersection over Union (IoU). As shown in Table~\ref{tab:ablation}, our approach achieved the best performance.

\begin{figure}[t!]
  \centering
  \includegraphics[width=0.9\linewidth]{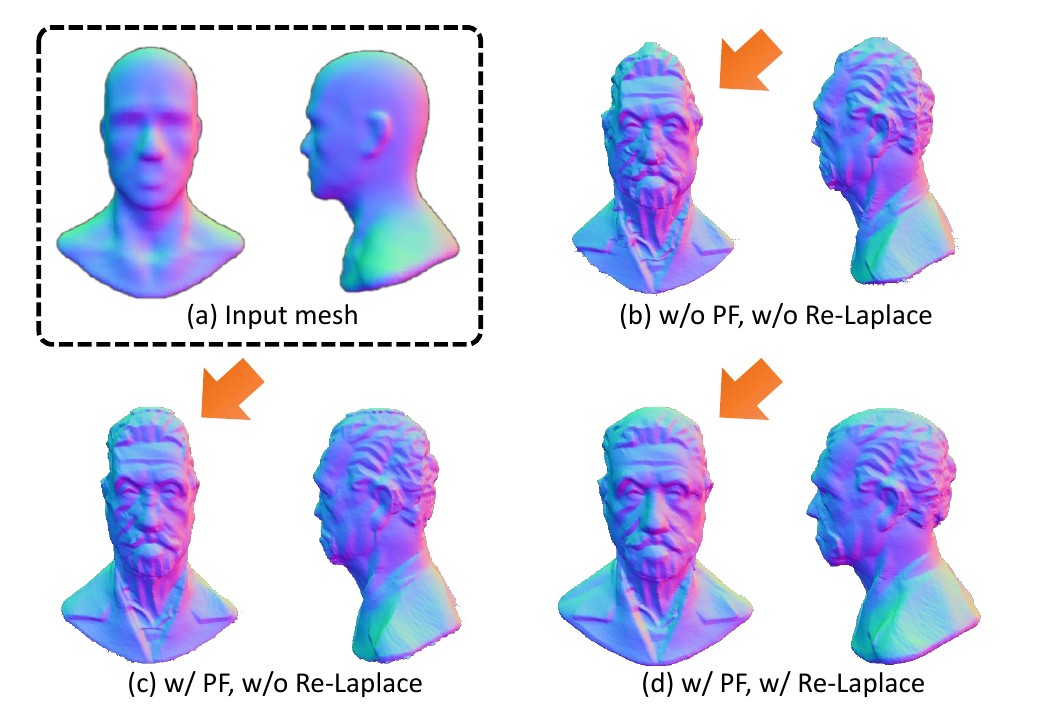}
  \caption{
  Ablation Study of the normal-based geometry refinement.
    We demonstrate the enhancement of our Poisson Fusion(PF) and Relative Laplace(Re-Laplace) module.
}
  \label{fig:ablation}
\vspace{-15 px}
\end{figure}

\textit{Regularizations During Mesh Optimization.} 
Our proposed regularization terms eliminate the
global distortions introduced in the detail enhancement process by normal stable diffusion, constraint the vertices towards the proximity of the coarse mesh, avoiding the mesh 
shrink introduced by the shape independent local smoothness
thereby enabling a robust optimization process.
As shown in Figure~\ref{fig:ablation}, directly refining the mesh without Poisson Fusion (PF) and Relative Laplace regularization (R-Laplace) results in an oddly sharp head due to global bias from normal stable diffusion. 
Although Poisson Fusion corrects this bias, the shape still shrinks. 
Replacing Original Laplace regularization with R-Laplace leading to a more reasonable shape.

\section{Conclusion and Discussion}

We present \textit{CraftsMan3D}, a pioneering framework for the creation of high-fidelity 3D meshes that mimics the modeling process of a craftsman, all within a mere 30 seconds. 
Our approach begins with the generation of a coarse geometry, followed by a refinement phase that enhances surface details. 
Despite our method's capability 
to produce high-quality 3D meshes,
the controllability of the Latent Set Diffusion model warrants further investigation, and the generation of texture for 3D meshes presents a promising avenue for future research.

{
    \small
    \bibliographystyle{ieeenat_fullname}
    \bibliography{bibliography}
}

\end{document}